# Effects of dynamic capability and marketing strategy on the organizational performance of the banking sector in Makassar, Indonesia

Efectos de las capacidades dinámicas y las estrategias de mercadeo en el desempeño organizacional del sector bancario en Makassar, Indonesia

MUHAMMADIN, Akhmad [1]
RAMLI, Rashila [2]
RIDJAL, Syamsul [3]
KANTO, Muhlis [4]
ALAM, Syamsul [5]
IDRIS, Hamzah [6]

**ABSTRACT**
The dynamic capability and marketing strategy are challenges to the banking sector in Indonesia. This study uses a survey method solving 39 banks in Makassar. Data collection was conducted of questionnaires. The results show that, the dynamic capability has a positive yet insignificant impact on the organizational performance, the marketing strategy has a positive and significant effect on organizational performance and, dynamic capability and marketing strategy have a positive and significant effect on the organization's performance in the banking sector in Makassar.

**RESUMEN**
Sustituir la versión en español del Resumen: La capacidad dinámica y la estrategia de marketing son desafíos para el sector bancario en Makassar. Este estudio utiliza un método de encuesta que resuelve 39 bancos en Makassar. La recolección de datos se realizó de cuestionarios. Los resultados muestran que, la capacidad dinámica tiene un impacto positivo pero insignificante en el rendimiento de la organización, la estrategia de marketing tiene un efecto positivo y significativo en el rendimiento de la organización y, la capacidad dinámica y la estrategia de marketing tienen un efecto positivo y significativo en el rendimiento de la organización en el sector bancario en Makassar.
**Palabras clave:** capacidad dinámica, estrategia de mercadeo, desempeño organizacional, banca

[1] Postgradute, Makassar Economic College (STIEM-Bongaya), Makassar, Indonesia, and 2. Department of Institute of Malaysia and International Studies (IKMAS) Universiti Kebangsaan Malaysia, Selangor Malaysia. E-mail : muhammadin.akhmad@yahoo.com
[2] Department of Institute of Malaysia and International Studies (IKMAS) Universiti Kebangsaan Malaysia, Selangor Malaysia. E-mail : rashramli@yahoo.com
[3] Postgradute, Makassar Economic College (STIEM-Bongaya), Makassar, Indonesia. E-mail : ridjalsyamsul@yahoo.com
[4] Postgradute, Makassar Economic College (STIEM-Bongaya), Makassar, Indonesia. E-mail : mukliskanto1@gmail.com
[5] Postgradute, Makassar Economic College (STIEM-Bongaya), Makassar, Indonesia. E-mail : syamsul.alam2010@gmail.com
[6] Makassar Economic College (STIEM-Bongaya), Makassar, Indonesia. E-mail : hamzah.idris.1980@gmail.com

# 1. Introduction

Strategic management is a process that is part of a manager's job to manage a complex organization and to apply strategies in establishing relationships between organizational efficiency and organizational opportunities (Mishina et al. 2004).

Management that leads to business opportunities that is, how managers can gather and mobilize the various resources is needed to capitalize on the opportunities Hitt et al. (2013). The most widely discussed question in the field of strategic management is why one company is more successful than another and how to make the company more successful (Porter 2012; Mishina et al. 2004). This problem can be formulated as a link between the internal environment which is a very basic and interesting strategy to study in depth in the field of strategic management.

In this regard, banks as financial institutions face similar challenges and demands. Internal environments such as resources capabilities and core competencies can influence economic pressures both within the company and outside the firm itself which influence bank performance (Hitt et al. 2013). Demand of the ability to strive while still following the rules of banking to improve performance is a must. However, to recognize that competition between local and foreign banks influences the domestic economy and the need to implement prudent rules such as risk management and good corporate governance must be implemented (Teece et al. 1997). Here, the bank becomes a "special" institution that must coordinate all of the problems or must survive or be profitable.

In the banking industry there is uncertainty arising from the work environment as in other industries (Porter 2004; Pearce and Robinson 2013). The ability to adapt to the environment apart from being influenced by strategic capabilities is also influenced by strategic choices which further determine the performance (Pearce and Robinson (2013).

There are 120 entities in Indonesian banking industry with a very diverse capital ranging from of IDR 2.4 trillion to IDR 28.4 trillion, as equity per December 2008. Currently, there are significant changes in the environment resulting in more difficult forecasts competition and tightening in the field of banking (Booklet Banking Indonesia 2007). The environment is increasingly difficult to predict given the globalization and integration of Indonesia's economy with the world economy. Banking competition increases due to several factors. They include Indonesia's commitment to institutions, or to international and regional forums such as the World Trade Organization, ASEAN Free Trade Area, ASEAN Framework Agrement on Services and ASEAN Economic Community as well as globalization and integration of the financial sector in Indonesia which has greatly influenced the increasing number of foreign banks entering Indonesia through the creation of a joint venture bank with the domestic or through the acquisition (Bank Indonesia Annual Report, 2008).

**Table 1**
Bank ownership market, based on total assets

| Owner | 2016 | | | 2017 | | | 2018 | | |
|---|---|---|---|---|---|---|---|---|---|
| | Total Number of Bank | Total Number of Asset (IDR) | Segment | Total Number of Bank | Total Number of Asset (IDR) | Segment | Total Number of Bank | Total Number of Asset (IDR) | Segment |
| Goverment Bank | 30 | 2.147 | 45.0% | 30 | 2.516 | 46.5% | 30 | 2.788 | 32.9% |
| Foreign Exchange Bank | 36 | 1.818 | 38.1% | 38 | 1.999 | 37.0% | 40 | 2.243 | 26.6% |
| Non Foreign Exchange Bank | 30 | 128 | 2.7% | 29 | 140 | 2.6% | 27 | 2.617 | 30.9% |

| Owner | 2016 | | | 2017 | | | 2018 | | |
|---|---|---|---|---|---|---|---|---|---|
| | Total Number of Bank | Total Number of Asset (IDR) | Segment | Total Number of Bank | Total Number of Asset (IDR) | Segment | Total Number of Bank | Total Number of Asset (IDR) | Segment |
| Mix Bank | 14 | 287 | 6.0% | 12 | 320 | 5.9% | 11 | 343 | 4.0% |
| Foreign Bank | 10 | 390 | 8.2% | 10 | 432 | 8.0% | 10 | 473 | 5.6% |
| Total | 120 | 4.770 | 100% | 119 | 5.407 | 100% | 118 | 8.464 | 100% |

Source: Bank Indonesia Annual Report (2017)

Based on ownership, between 2016 and 2018, the number of government and foreign ownership of banking in Indonesia was relatively flat. This indicates that the increasing role of foreigners in the Indonesian banking industry will certainly affect the level of competition in the market. Although there are 120 commercial banks, the structure of Indonesia's banking industry focused on only a few banks. Based on the concentration ratio, there are currently fifteen leading banks the majority (71%) of Bank Indonesia's total banking assets (2018).

**Figure 1**
Levels of Banking Economic Development in Indonesia

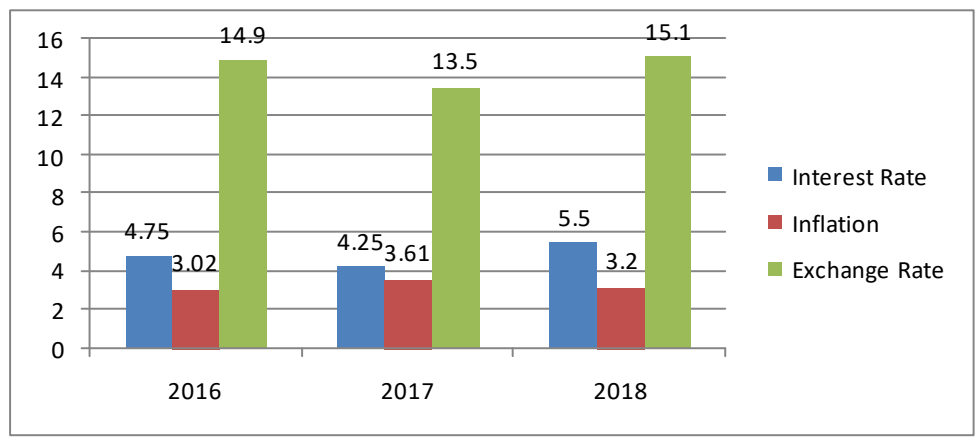

At the heart of Indonesia's banking structure above, over the course of three years there have been changes affecting the banking industry, namely interest rate fluctuations and an increase in the IDR against the US Dollar thus affecting the banking industry environment and tightening of banking regulations in Indonesia. Its influence is due to globalization influences such as the subprime morgage crisis in the United States, in 2015 and the global crisis caused by the Lira exchange rate crisis that plagued Turkey and Argentina's financial crisis in 2018.

## 1.1. Research problem

Given the global economic conditions, the banking industry including Indonesia has undergone a change and dynamic in recent years. Amid the changes in Indonesia's banking structure over the past 10 years there have been two changes affecting the banking industry namely the changing banking industry environment and tightening of banking regulations in Indonesia. Perceived from the internal environment, the industrial environment experienced a drastic change from the general environment in 2015, mainly due to the effects of globalization such as the crisis in Greece which has given a worldwide picture including in Indonesia (Bank Indonesia Annual Report 2015).

As described above, the banking industry including in Indonesia has faced the challenge of changing the internal environment in recent years. The banking industry is also regulated by prudent regulation at the heart of the market structure which is focused on several large banks both in terms of funds assets and credit regulations. In connection with this the main issues examined are the factors that lead to differences in banking performance which are associated with competitive advantage strategies (Porter 1980,1985,2004).

In addition, modern marketing theory states that the key to a competitive success of companies depends on the dynamic adaptation in a complex and ever-changing environment. These changes are characterized by an increase in sales, products or services due to fluctuations in external factors (politics, social economy, technology, law and the environment) (Pearce and Robinson's (2013) internal factors, resources, marketing, management, capabilities, efficiency and motivation) Hitt et al. (2013).

Banking problems in Makassar, which undergo significant growth, are marked by an increasing number of the opening of regional headquarters and new unit branches, showing an increasing trend of profits every year such as Mandiri Bank, BRI Bank, BCA Bank, BNI Bank, CIMB Niaga Bank, Danamon Bank, Permata Bank, BTN Bank, Maybank Indonesia and Hasa Mitra Bank which enjoy their assets grow each year. However, there are some banks that have shown volatile, unpredictable developments such as Maspion Bank, Mayapada Bank, Sinar Mas Bank, etc., while some banks have shown declining achievements such as Pataru Laba Bank which resulting in closure (revocation of operating license) such as the Niaga Madani Community Credit Bank.

## 1.2. Research questions

In the same industrial environment, why some banks are experiencing rapid growth while others are declining instead. Theoretically and empirically there is no conclusive opinion to answer the problem. Therefore, interesting to investigate are the factors of dynamic capabilities, marketing strategies towards organizational achievement which can provide satisfaction and loyalty in financial institutions such as banking. Based on the subject matter above the research questions you want to answer, so this study can be studied in three reasons,

1. What is the dynamic capability influence to the organizational performance in the banking sector in Makassar?

2. How do marketing strategy influence to the organizational performance in the banking sector in Makassar?

3. Why are dynamic capability and marketing strategy influence to the organizational performance in the banking sector in Makassar?

## 1.3. objectives of the research

The objective of this study is to analyze the influence of dynamic capability marketing strategy on the organizational performance of banking sector in Makassar. The objectives of this study are:

1. To test the impact of dynamic capability, on organizational performance on the banking sector in Makassar.

2. To analyze the impact of marketing strategy, on organizational performance on the banking sector in Makassar.

3. To identify the impact of dynamic capability, and marketing strategy, on organizational performance on the banking sector in Makassar.

## 2. literature review

### 2.1. dynamic capability

Dynamic capability is the ability of a team's resources to carry out task or activity (Segal-Horn's, 2004). This capability enables the company to exploit its resources in implementing its strategy Hitt et al. (2014). In general, organizations have many capabilities but Marcus's (2005) limited number of competencies. Competence will enable managers to combine resources with the ability in transforming and serving to meet customer needs. Together with the resources and capabilities, the organization creates distinctive competence (Marcus, 2005). There are several models that show the relationship between resources and capabilities as determinants of strategic capabilities. As shown in the following diagram resources need to be integrated with the ability to generate the capabilities of strategy (Hubbard, 2000). Capability is the process system or community of the organization to coordinate the resources to be used productively by the company.

Therefore, the above description shows that expert opinion from several findings shows that in order to create effective strategies, supports by the resources and the ability as well as other competencies are mandatory to achieve positive excellence. Businesses choose specific strategies based on internal strengths (effectiveness) and environmental conditions (DeSarbo et al. 2005). The relationship of market orientation with business performance through the factor of innovation capability is encouraged (Patteri et al. 2014). This is also in line with the opinion of Akhmad Muhammadin and Rashila Ramli (2018) the environment has a positive effect on financial performance in the banking industry in Indonesia.

Furthermore, Teece (2009) emphasizes that a company's dynamic capabilities describes how an organization develops its capabilities from a range of competencies in a dynamic business environment. Some dimensions of capability which need to be built on are (i) organizational learning capabilities, (ii) relationship capabilities, and (iii) management capabilities. Organizational learning capabilities play a role in enhancing economic value-added through the development of new ideas and renewing existing corporate capabilities (Teece 2009). In companies, learning processes include individual, group and organizational skills (Zollo and Winter 2002)

Companies can also collaborate and partner with other parties including competitors for new organizational learning, helping companies to identify internal weaknesses which in turn will have a positive impact on performance (Moliterno, 2007). Zott (2003) found that change in the capabilities of the company affecting performance. Similarly, Lenox and King (2004) have found that by developing the power of dynamic capabilities, companies are more successful and perform better than other companies. Based on the above, the following hypothesis is formulated,

> **H1**: Dynamic capabilities have a positive impact to the organizational performance.

## 2.2. Marketing strategy

Marketing is a combination of interrelated activities to understand consumer needs and to develop promotions, distributions, services and prices so that consumers' needs are met to a certain extent (Kotler, 2012). With the advent of consumer marketing, it is no longer necessary to meet individual needs by exchanging between users and marketers so that there will be a large number of users for the activities to be handled. According to Kotler (2012), marketing is an organizational function and a set of processes for creating communicating, and delivering value to customers and managing customer relationships in ways that benefit the organization and stakeholders.

The easiest way to observe and analyze company performance strategies is through structural analysis. Structural resources not only involve financial physical and human resources in each field but also the ability of personnel in each field to formulate and execute organizational goals, strategies and policies. When used properly these resources can serve as a strength of the company to support strategic decisions in achieving the superior performance (Wheelen and Hunger, 2007). The banking sector requires appropriate strategies by implementing a generic strategy to improve the performance of the organization Syamsul Ridjal and Akhmad Muhammadin (2018)

Competitiveness strategies are a process of strategic planning reviewing a company's marketing and distribution research and development, production and operations, human and employee resources as well as financial and accounting factors to determine where a company has key capabilities so that the company leverages opportunities in the best way (Kotler, 2012). Therefore, to find out the influence of strategy in the company, the functional analysis of the company is employed which includes marketing functions.

**H2**: Marketing strategies have a positive impact on organizational performance

## 2.3. Organizational performance

Strategy management is the decision and process that managers use to make long-term adaptation of the company to its environment. Managers need to manage their business for long terms as well as to refine their means of production and development in research to keep the company competitive and to meet changing customer needs (Simon et al. 2000). Managers cannot handle well what is not measured (Kaplan and Norton 2004). Therefore, accurate measurement of strategy performance is a measure that assesses the superiority or quality of a company's long-term adaptation (Berger et al. 2009).

Another study that looked at performance factors was the study of Thomas et al. (2008) who suggest that company performance is overall performance in terms of company quality in the context of product quality, service, employee morale, employee expertise, productivity, workforce efficiency and profitability. The component is referred to as an effort to increase profits from the sales revenue of products and services.

Kotler (2013) view that quality is the ability of a product to perform its basic function depending on the quality produced. Li et al. (2006) argue that companies have a competitive advantage when they are able to provide quality product and reliable organizational performance which can provide higher value to customers. Companies with high quality products will be able to set specific prices which in turn will increase their profit margins or return on investment (Li et al. 2006).

Organizational performance by making adjustments based on the conditions in the organization (Kumar and Strandholm 2002). This is in support of a previous study conducted by Koo et al. (2004) that strategy differences influence the performance of companies. The findings of this study indicate that, the variables of strategic difference are significantly related to the firm's performance variables.

According to the Bank Indonesia Report (2008, 2013), by the end of 2013 the banking industry was relatively stable. Most or 82.4% of the public banks were in supervisory status, with a significant improvement over the previous year. This is because, as banks were previously listed as under poor supervision, they have been able to be upgraded to banks under normal supervision. Bank-related issues can be resolved in accordance with a plan of action which has been agreed between Bank Indonesia (BI) and the bank's owner or manager. Meanwhile, there are banks that are still under scrutiny and under special supervision. This is because as at the end of 2013 bank managers and owners have not been able to resolve the issues and have not complied with the bank's commitments as expressed in the bank's stability action plan (Publisher Banking Supervision Report BI 2008, 2013).

**H3:** Dynamic capability and marketing strategies have a positive impact on organizational performance.

## 3. Methodology

This study was conducted through primary and secondary data collection. Primary data were obtained by distributing the questionnaire to the respondents who met certain criteria. The questionnaire contained a number of questions with sufficient explanations to facilitate the respondents in filling them out. This study was also conducted using a library research approach with secondary data. Secondary data were obtained from the Indonesian bank's annual reports, Indonesian economic reports, books, scholarly papers and proceedings in an effort to strengthen the basic concepts of analysis.

### 3.1. Population, sample and selection of respondents

The unit population of this study is the whole commercial public bank located in Makassar which is a total of 67 banks. This study used survey methods to encourage questions for more reliable information. This is required as the data analysis model to be used is the Statistical Package for the Social Sciences (SPSS). The unit of analysis is each bank with a target of respondents consisting of branch managers, marketing managers and operations managers or branch leaders one level below the authorized manager to answer the questionnaire. The branch manager or leader is considered to have deep knowledge of the strategies and policies of the commercial banks they lead.

**Table 2**
Total Population and Sample

| No | Type of Bank | Total Participation | Answer | Participation No Answer |
|----|--------------|---------------------|--------|-------------------------|
| 1  | Government Bank | 5 | 5 | - |
| 2  | National Private Bank | 42 | 23 | 19 |
| 4  | Rural Bank | 13 | 7 | 6 |
| 5  | Mix Bank | 3 | 3 | - |
| 6  | Foreign Bank | 4 | 1 | 3 |
|    | Total | 67 | 39 | 28 |

Of the total population of 67 banks located in Makassar as of the end of October 2018, 39 banks (58.21%) participated in this study which was considered to be a sufficient number to be analyzed. Of these 39 sets of valid questionnaires were used for further processing.

## 4. The results of data analysis

A total of 67 sets of questionnaires were sent directly to all banks in the population (1 questionnaire per bank). Of the total questionnaires distributed, 39 banks returned the questionnaire (a response rate of 58.2%) and after reviewing the questionnaire were all completed and considered valid for processing. So the respondents who filled out the questionnaire consisted of 5 government banks, 23 national private banks, 7 community credit banks, 3 mixed banks, and 1 foreign bank.

### 4.1. Response description based on dynamic capabilities

The response of the respondents regarding their ability to influence the performance of the organization the banking sector in Makassar to determine the ability of bank management as compared to other major competitors can be seen in table 3 below.

From the respondents' perceptions of capacity, the majority of respondents' responses were strongly disagree, agree and strongly agree, as can be seen from, the first statement of respondents gave an answer of 17, or 43.6%. In the second statement, respondents answered yes were 18 or 46.2%. For the third statement, 19 or 48.7% of respondents gave an agreeing response. The fourth statement was given agreement by 19 or 48.7% respondents. The fifth statement was agreed by 17 or 43.6% of samples. For the sixth respondents, agreements were given by 14 respondents, or 35.9%. In the seventh statement, respondents answered yes were 17 or 43.6%. The eighth statement was given agreement by 21, or 53.8% respondent. The ninth statement was given agreement by 19, or 48.7% respondent. The tenth statement was the most agreeable answer, agreed by 17 or 43.6% respondent.

**Table 3**
Respondent's Description on Dynamic Capability

| Indicators | strongly disagree | do not agree | disagree | agree | strongly agree |
|---|---|---|---|---|---|
| Our bank has the flexibility to communicate and coordinate effectively across teams / divisions / posts | - | - | 8<br>20,5 | 17<br>43,6 | 14<br>35,9 |
| Our bank has the ability to quickly identify new business opportunities or potential threats to existing businesses | - | - | 9<br>23,1 | 18<br>46,2 | 12<br>30,8 |
| Our bank helps employees to maintain a balanced life between working with family | - | - | 10<br>25,6 | 19<br>48,7 | 10<br>25,6 |
| Our bank has the flexibility to understand the customer's specific needs | - | 1<br>2,6 | 10<br>25,6 | 19<br>48,7 | 9<br>23,1 |
| Our bank has the ability to secure and retain top managers who have good and reliable relationships | - | - | 12<br>30,8 | 17<br>43,6 | 10<br>25,6 |
| Our bank has the capability to control the overall performance of the organization | - | - | 9<br>23,1 | 14<br>35,9 | 16<br>41,0 |
| Our bank has the ability to interact and co-operate with the community in the context of mutually beneficial relationships | - | - | 7<br>17,9 | 15<br>38,5 | 17<br>43,6 |
| Our bank has the capability to develop and communicate an organizational vision which is understood by all members of the organization | - | 1<br>2,6 | 8<br>20,5 | 21<br>53,8 | 9<br>23,1 |
| Our bank has the ability to unite different views improve coordination and cooperation between key executives to provide better organizational performance | - | - | 10<br>25,6 | 19<br>48,7 | 10<br>25,6 |
| Our bank has the capability to enhance participation in decision making, at the middle and upper management levels | - | - | 7<br>17,9 | 15<br>38,5 | 17<br>43,6 |

## 4.2. Description of respondents, according to the marketing strategy

The response of the respondents to the marketing strategy in relation to the influence of the organization the banking sector in Makassar in deciding on the choice of bank strategy last year compared to the other major competitors can be seen in table 4 as follows

**Table 4**
Description of Respondents, On Marketing Strategy

| Indicators | strongly disagree | do not agree | disagree | agree | strongly agree |
|---|---|---|---|---|---|
| Our company's reputation including image and reputation in the eyes of our customers is related to our product marketing business | - | - | 10<br>25,6 | 10<br>25,6 | 19<br>48,7 |
| Customers buy products paying attention to the quality (quality) of our products | - | - | 6<br>15,4 | 16<br>41,0 | 17<br>43,6 |
| Our attitude and our way, to serve our customers, is satisfactory | - | 2<br>5,1 | 7<br>17,9 | 13<br>33,3 | 17<br>43,6 |
| A variety of products are offered so that customers can choose and decide on the product according to their needs and requirements | - | - | 13<br>33,3 | 8<br>20,5 | 18<br>46,2 |
| We are constantly innovating to refine and produce better new products to appeal to customers | - | 1<br>2,6 | 8<br>20,5 | 13<br>33,3 | 17<br>43,6 |
| Our office space can reach customers | - | - | 6<br>15,4 | 20<br>51,3 | 13<br>33,3 |
| Efforts to communicate or introduce products to customers | - | - | 8<br>20.5 | 21<br>53,8 | 10<br>25,6 |
| We are effectively promoting the product | - | - | 13<br>33,3 | 21<br>53,8 | 5<br>12,8 |

On respondents perception of marketing, most of the answers were agree and strongly agree which can be seen from the first statement where respondents gave a very agreeable answer from 19 peoples or 48.7%. Seventeen or 43.6% of the respondent gave a very agreeable response for the second statement. The third respondent got very agreeable responses from 17 or 43.6% of respondents. The fourth statement is very agreeable by 18 peoples or 46.2%. The fifth statement is answered as very agreeable by17 peoples or 43.6%. The sixth statement was given a consensus response from 20 peoples or 51.3%. Respondents gave an answer of 21 peoples or 53.8% for the seventh statement. The eighth statement is answered by 21 peoples or 53.8% of respondents.

### 4.3. Response description, based on organizational performance (quality)

In terms of quality one of the notable variables as a dependent variable is the determination of the banks strategic choices compared to other main competitors, which can be seen in the follow table 5

**Table 5**
Responses on organizational quality achievements

| Indicators | Respondent Response | | | | |
| --- | --- | --- | --- | --- | --- |
| | strongly disagree | do not agree | Disagreed | agree | strongly agree |
| Last year compared to our main competitors the change in our bank's competitive advantage was much better | - | 5<br>12,8 | 8<br>20,5 | 17<br>43,6 | 9<br>23,1 |
| Last year compared to our major competitors the relative change in our bank lending rates was much better | 1<br>2,6 | 5<br>12,8 | 11<br>28,2 | 11<br>28,2 | 11<br>28,2 |
| Last year compared to our major competitors the change in the growth of our bank lending rates was much better | - | 3<br>7,7 | 8<br>20,5 | 6<br>41,0 | 12<br>30,8 |
| Last year compared to our main competitors our third-party/bank division was much better | - | 1<br>2,6 | 8<br>20,5 | 15<br>38,5 | 15<br>38,5 |
| Last year compared to our main competitors the level of satisfaction of our depositors and third-party funds/bank communities was much better. | - | 4<br>10,3 | 10<br>25,6 | 12<br>30,8 | 13<br>33,3 |

Based on the respondents 'perception of quality the majority of respondents' responses were strongly agreed and strongly agree as can be seen from the first statement where respondents gave an answer of 17 peoples or 43.6%. In the second statement respondents gave their respective disagreements, agree and strongly agree with a total of 11 peoples or 28.2%. In the third statement the respondents gave a consensus response of 16 peoples or 41.0%. In the fourth statement the respondents gave a very agreeable and agreeable answer each responded by 15 peoples or 38.5%. On fifth statement respondents gave a very agreeable answer coming from 13 peoples or 33.3%.

### 4.4. Data validity test

Validity tests are used to test the accuracy of the measuring instrument to reveal the concept of the indication or event being measured. The constructive validity in this study was tested using bivariate person (product moment correlation). Test results using the SPSS version 22 program are valid when they have a correlation value of product moment exceeding 0.40.

**Table 6**
Validity Test

|  | | | | |
|---|---|---|---|---|
| Dynamic Capability | X2.1 | 0,641 | 0,40 | Valid |
|  | X2.2 | 0,507 | 0,40 | Valid |
|  | X2.3 | 0,517 | 0,40 | Valid |
|  | X2.4 | 0,595 | 0,40 | Valid |
|  | X2.5 | 0,416 | 0,40 | Valid |
|  | X2.6 | 0,647 | 0,40 | Valid |
|  | X2.7 | 0,488 | 0,40 | Valid |
|  | X2.8 | 0,459 | 0,40 | Valid |
|  | X2.9 | 0,494 | 0,40 | Valid |
|  | X2.10 | 0,579 | 0,40 | valid |
| Marketing Strategy | X4.1 | 0,639 | 0,40 | Valid |
|  | X4.2 | 0,568 | 0,40 | Valid |
|  | X4.3 | 0,766 | 0,40 | Valid |
|  | X4.4 | 0,758 | 0,40 | Valid |
|  | X4.5 | 0,855 | 0,40 | Valid |
|  | X4.6 | 0,509 | 0,40 | Valid |
|  | X4.7 | 0,460 | 0,40 | Valid |
|  | X4.8 | 0,413 | 0,40 | Valid |
| Quality | Y9.1 | 0,699 | 0,40 | Valid |
|  | Y9.2 | 0,809 | 0,40 | Valid |
|  | Y9.3 | 0,758 | 0,40 | Valid |
|  | Y9.4 | 0,624 | 0,40 | Valid |
|  | Y9.5 | 0,766 | 0,40 | Valid |

## 4.5. Reliability test

Reliability test is used to determine whether the indicator or questionnaire used is reliable as a variable measuring instrument. The reliability of an indicator or questionnaire can be derived from the Cronbach's alpha (α) value greater than 0.60, therefore the indicator or questionnaire is reliable, whereas when the Cronbach's alpha (α) is smaller than 0.60 then the proposed indicator or questionnaire is not reliable. The overall reliability test results can be found in the following table:

**Table 7**
Reliability test

| | | | | |
|---|---|---|---|---|
| Dynamic Capability | 10 | 0.860 | 0,60 | Accepted |
| Marketing Strategy | 8 | 0.817 | 0,60 | Accepted |
| Quality | 5 | 0.785 | 0,60 | Accepted |

## 4.6. Multicollinearity test

The multicollinearity test aims to determine whether there are independent variables that have similarities between the independent variables in the regression model. In the event of a correlation it is stated that the regression model has multicollinearity problems. Multicollinearity tests were performed taking into account Variance Inflation Factor (VIF) tolerance values. If tolerance values exceed 0.10 and VIF <10, then no multicollinearity is indicated. The results of the multicollinearity test can be seen in the following table

**Table 8**
Multicollinearity test

| Variables | Collinearity Statistics | | VIF Standard | Decision |
|---|---|---|---|---|
| | Tolerance | VIF | | |
| Dynamic Capability | 1,000 | 1,000 | 10 | No multicollinearity |
| Marketing Strategy | 1,000 | 1,000 | 10 | No multicollinearity |

### 4.7. Heteroscedastisity test

The Heteroscedastisity test aims to test whether a regression model has variance differences from one observation to another. Linear test is used to detect heteroscedastisity in this study. This test compares the significant of the test to the standard value. If it is significant of <0.05, it can be concluded that the regression model contains heteroscedastisity otherwise the significance value is> 0.05, then heteroscedastisity occurs. The results of the heteroscedastisity test can be seen in the following table:

**Table 9**
Heteroscedastisity test

| No. | Variables | Significant | Description |
|---|---|---|---|
| 1. | Dynamic Capability | 0,246 | no Heteroscedastisity |
| 2. | Marketing Strategy | 0,052 | no Heteroscedastisity |

### 4.8. Normality test

Normality tests are intended to test whether regression is variable or has a normal distribution. Normal tests can be performed using the One Sample Kolmogorov-Smirnov Test with a significance level of 0.05 or 5%. If significance is>0.05, data distribution is normalized. On the other hand, if significance <0.05 was generated, data would not be normally distributed. The results of the Kolmogorov-Smirnov test values, for the model obtained, can be seen in table 9 below:

**Table 10**
Normalization Test with One Sample Kolmogorov-Smirnov Test

| | | Dynamic Capability and Marketing Strategy |
|---|---|---|
| N | | 39 |
| Normal Parameters[a,b] | Mean | 99.33 |
| | Std. Deviation | 8.514 |
| Most Extreme Differences | Absolute | .086 |
| | Positive | .044 |
| | Negative | -.086 |
| Test Statistic | | .086 |
| Asymp. Sig. (2-tailed) | | .200[c,d] |

a. Test distribution is Normal.
b. Calculated from data.
c. Lilliefors Significance Correction.
d. This is a lower bound of the true significance.

| | | Organizational Performance |
|---|---|---|
| N | | 39 |
| Normal Parameters[a,b] | Mean | 38.56 |
| | Std. Deviation | 5.924 |

| Most Extreme Differences | Absolute | .132 |
|---|---|---|
| | Positive | .073 |
| | Negative | -.132 |
| Test Statistic | | .132 |
| Asymp. Sig. (2-tailed) | | .085[c] |

a. Test distribution is Normal.

b. Calculated from data.

c. Lilliefors Significance Correction.

### 4.9. Multiple linear regression analysis

Multiple linear regression analyzes were used to determine whether the influence of dynamic capabilities, as independent variables and marketing strategies, on dependent variables (cost performance and quality). The results of the regression analysis can be seen in the following table:

**Table 11**
Results of regression equation

| Model | Unstandardized Coefficients | | Standardized Coefficients | T | Sig. |
|---|---|---|---|---|---|
| | B | Std. Error | Beta | | |
| (Constant) | 30,973 | 13,441 | | 2.304 | 0,027 |
| Dynamic Capability | 0,152 | 0,090 | -0,263 | -1,693 | 0,099 |
| Marketing Strategy | 0,255 | 0,108 | 0,366 | 2,353 | 0,024 |

Based on the results above, when expressed in standardized coefficient, the regression equation is as follows

$$Y = \alpha + \beta_1 X_1 + \beta_2 X_2 + \acute{\epsilon}$$

$$Y = 30,973 + 0,152 X_1 + 0,255 X_2$$

### 4.10. Analysis of correlation and coefficient of determination

Correlation coefficients are used to see the relation of independent variables in describing bound variables, while the coefficient of determination ($R^2$) is the basis of measurement of the extent to which the model is able to explain the variation of dependent variables. The coefficient of determination is between 0 and 1.

The coefficient of determination ($R^2$) is small, which means that the independent variable to explain the bounded variable is very limited. A value close to 1 means that the independent variable provides all the information it needs, to predict the variation in the bound variable. The magnitude of the coefficient of determination can be found by the squared value of the correlation coefficient. The results of the calculation coefficients of the two regression models used in this study are as follows:

**Table 12**
Model Summary

| Model | R | R Square | Adjusted R Square | Std. Error of the Estimate |
|---|---|---|---|---|
| 1 | 0,406[a] | 0,165 | 0,118 | 5,563 |

a. Predictors: (Constant), Marketing Strategy, Dynamic Capability

### 4.11. Individual parameter test (t-test)

This test is used to prove significant influence, between independent variables (dynamic capability and marketing strategy) on dependent variables (organizational performance), which can be done by comparing standard values if the value of ρ is smaller than the standard value. This means that the hypothesis is presented or has significant influence. Conversely, if the value of ρ is greater than the standard value then it is insignificant or the hypothesis is rejected.

**Table 13**
Partial results of the tests (t-test)

| Model | ρ-value | Standard Value |
|---|---|---|
| Dynamic Capability | 0,099 | 0,05 |
| Marketing Strategy | 0,024 | 0,05 |

## 4.12. Simultaneous parameter test (t-test)

The t-test in this study using ANOVA test is to determine the effect of simultaneous variables (dynamic capabilities and marketing strategies) on dependent variables (organizational performance). The basis for making a decision is that if significance is less than 0.05 then Ho is rejected and otherwise Ha is accepted. This means that the simultaneous independent variables have a simultaneous effect on the dependent variable. If the significance value is greater than 0.05 then Ho is accepted and Ha rejected. This means that the independent variables have no simultaneous effect on the dependent variables. The results of the t-test are presented in the following table.

**Table 14**
ANOVA test

| Model | | Sum of Squares | Df | Mean Square | F | Sig. |
|---|---|---|---|---|---|---|
| 1 | Regression | 219,698 | 2 | 109,849 | 3,550 | 0,039[b] |
|   | Residual | 1113,892 | 36 | 30,941 | | |
|   | Total | 1333,590 | 38 | | | |

a. Dependent Variable: Organizational achievements
b. Predictors: (Constant), competitive advantages, internal environment

## 4.13. Influence of the internal environment, on organizational performance

Based on the results of the regression equation between the dynamic and organizational capabilities of the banking sector in the city of Makassar, the regression coefficient is 0.152 and the significance value is 0.099 having a probability of 0.000 <0.05 which means that it has positive and non-significant impact on the performance of the organization in the banking sector in Makassar.

Based on the results of the questionnaire it was found that the positive and negative dynamic ability on organizational performance as the majority of respondents answered that dynamic ability affects performance in the organization and has no significant effect. The dynamic capability component (hypothesis H1) of the source analysis obtained a value of 0.152 that the direct effect of the dynamic ability variable on organizational performance variables was 15.2%. This means that dynamic capability variables have a direct and positive influence but are not significant in measuring the performance variables of the organization the banking sector with the contribution of factors, capital, departmental network and reputation to the organization's performance by 15%.

## 4.14. Influence of marketing strategies on organizational performance

From the results of the processed data regression equation, the regression coefficient is obtained for a marketing strategy variable of 0.255 and a probability value of 0.024. It can be concluded that the marketing strategy has a positive and significant impact on the performance of the banking sector organization in Makassar.

Marketing strategy factors have a positive relationship with organizational performance where marketing variables have a significant impact. Varying marketing as a factor in marketing strategy has a value of 25.5% thus the better the marketing the better the reputation of the company in the eyes of the customer. The attitude of the service and the volume of products offered will be able to meet the needs and desires of the customers and businesses to communicate or introduce the product in the form of promotion to attract customers.

## 4.15. Influence of dynamic capabilities and marketing strategies on organizational performance

Based on the results of the regression equation, the dynamic capabilities and marketing strategies obtained a regression coefficient of 0.255 which additionally has a probability value of 0.024 <0.05 which means that the product has a positive and significant influence on the performance of the banking sector organization in Makassar.

Based on at the results of the questionnaire distribution, the findings suggest that the internal environment has significant impact on organizational performance. Where the respondents' perception of dynamic capabilities is high it can influence the performance organization especially in the banking sector based in Makassar. Therefore, in supporting organizational performance it is important to note that dynamic capability and marketing strategy factors are maintained in the public eye therefore they can influence customers to remain loyal customers to organizational performance that includes cost and quality performance in the banking sector in Makassar.

## 5. Conclusion

Based on the problems of research, theories and hypotheses are developed, as well as through the test described in the research, it can be concluded as follows.

Dynamic capabilities have a positive and negative effect on organizational performance. This means that in order to improve bank performance in Makassar, it needs strength in a conducive corporate environment. Equally important is the choice of capabilities to execute well so that the indicators needed to be used in the banking sector in Makassar, namely capital, reputation, office networks, and ATM networks need to be enhanced simultaneously so that their influence can be felt broadly and evenly to unit or bank branch in Makassar. Therefore, the choice made by managers will be effective provided it is based on a comprehensive dynamic capability study. Likewise organizational performance will succeed if it is aligned with good performance.

Marketing strategies have a positive and significant impact on organizational performance. This means that the strategies set out in the banking business must be consistent with the organizational performance of the managers as decision makers. Marketing strategies towards good organizational performance, when supported by reliable marketing.

Dynamic capabilities and marketing strategies have a positive and significant impact on organizational performance. This means that to improve the performance of the banking sector in the city of Makassar, it should simultaneously pay close attention to marketing and implementing a comprehensive strategy as these two key indicators have a significant impact on the company's performance in banking practice. Ability to explore dynamic capabilities and marketing strategies can trigger trends in overall company performance orientation. It should be carried out consistently on the level of capability development that occurs in organizational performance.

## Bibliographic References